\documentstyle[twocolumn,prl,aps,epsf,floats]{revtex}

\begin{document}

\twocolumn[\hsize\textwidth\columnwidth\hsize\csname
@twocolumnfalse\endcsname

\title{ Electronic States in Two-Dimensional Triangular Cobalt Oxides: 
        Role of Electronic Correlation}
\author{Liang-Jian Zou, J.-L. Wang and Z. Zeng}
\address{\it Institute of Solid State Physics, Academia Sinica, 
             P. O. Box 1129, Hefei 230031, China}
\date{\today}
\maketitle

%%%%%%%%% %%%%%%%%% %%%%%%%%% %%%%%%%%% %%%%%%%%% %%%%%%%%% %%%%%%%%% 

\begin{abstract}
 
   We obtain the electronic states and structures of two-dimensional cobalt 
oxides, Na$_{x}$CoO$_{2}$ (x=0, 0.35, 0.5 and 0.75) by utilizing the 
full-potential linear muffin-tin orbitals (FP-LMTO) methods, from which 
some essential electronic interaction parameters are estimated: the bare
on-site Coulomb interaction of cobalt U$_{dd}$=7.5 eV renormalizes to 5
eV for x=0.35, the $pd$ hybridizations t$_{pd\sigma}$ and t$_{pd\pi}$ are
-1.40 and 0.70 eV, respectively. 
The density of states at E$_{F}$ decreases from 6-7 
states/eV in the local density approximation (LDA) to about 1.0 states/eV 
in the LDA+U scheme. The role of the intercalation of water molecules 
and the microscopic mechanism of the
superconductivity in Na$_{0.35}$CoO$_{2}$$\cdot$mH$_{2}$O is discussed.


\end{abstract}.

]

    The triangular cobalt oxides Na$_{x}$CoO$_{2}$ synthesized in recent 
years \cite{s1,s2} with large thermopower effect may be a hopeful candidate 
for the exotic resonating valence bond (RVB) phase \cite{s3}: well-defined 
local magnetic moment of Co$^{4+}$ indicates low spin S=$1/2$ \cite{s4}, 
highly anisotropic crystalline structure implies quasi-2-dimensional 
(2D) character \cite{s5}, and the triangular symmetry may lead to 
strong frustration effect of the antiferromagnetically coupled spins. 
However, the good metallic conduction
in Na$_{x}$CoO$_{2}$ seems against to the RVB phase, and its
thermoelectric coefficient is an order of magnitude larger than common
metals \cite{s2,s3,s4,s6,s7}, not expected for RVB metal. 
Two distinct interpretations were proposed to account for the origin of 
the unusual large thermopower effect: Singh \cite{s8}
suggested the large density of states (DOS) near the Fermi surface (FS)
results in the large thermoelectric effect; meanwhile, a theory of strongly
correlated interaction between 3d electrons \cite{s9,s10,s11} was proposed
to explain the large thermopower at room temperature. 
The main obstruction to settle this debate is that the electronic states 
of cobalt and oxygen are not well understood.

  Recently Takada {\it et al.} \cite{s12} reported a strange drop in
resistivity and a significant diamagnetic signal in magnetic susceptibility 
below an temperature T$_{p}$ =5 K in Na$_{0.35}$CoO$_{2}$$\cdot$mH$_{2}$O, in
which the water molecules were intercalated into the Na$^{+}$ and CoO$_{2}$
layers in Na$_{0.35}$CoO$_{2}$, these unusual phenomena are interpreted as the
onset of superconductivity (SC) below T$_{p}$. The diamagnetic
signature was confirmed by a more recent experiment \cite{s13}, and furthermore
an unambiguous zero-resistance state was found \cite{s14}, verifying the
the SC in hydrated 2D cobalt oxides. These
anomalous properties are distinct evidences that a novel
SC phase forms in Na$_{0.35}$CoO$_{2}$$\cdot$mH$_{2}$O below
T$_{p}$, the question thus arises what the nature of the novel
SC and its normal phases is. Doped 
cobalt oxides resemble to doped high T$_{C}$
cuprates in many aspects, for example, quasi-2D layer structure, 
Mott insulator becomes metal upon doping, the frustration effect,
etc. Baskaran \cite{s15} and Wang {\it et al.} \cite{s16} It was proposed 
that the Cooper pairing in SC phase originates from the
instability of RVB metallic background in 2D triangular cobalt oxides,
implying strongly correlated 3d electrons and the
exclusion of double occupation in 3d orbitals. At present our the
knowledge to some essential interaction parameters to model the physics 
in Na$_{0.35}$CoO$_{2}$ and Na$_{0.35}$CoO$_{2}$$\cdot$mH$_{2}$O, such as
the hopping integrals of the 3d electrons, $t$, and the Coulomb interaction 
between 3d electrons, $U$, etc,
are very poor, understanding the electronic states and structures is an
essential step to gain insight to the electronic properties and the role
of the intercalated water molecules in the novel SC and
normal phases, a good knowledge for Co 3d and O 2p band near Fermi suface
is required for any further reliable quantitative model.

   In this Paper we present the electronic state and structure results in 
Na$_{x}$CoO$_{2}$ (x=0, 0.35, 0.5 and 0.75) utilizing the first principle
full-potential linear Muffin-Tin orbitals (FP-LMTO) method and the local
density approximation (LDA)+U
method. The essential interaction parameters for further modeling the nature
of the new phase are estimated. We demonstrate that undoped CoO$_{2}$ is a 
quasi-2D intermediate correlated charge transfer insulator, Na-doped  
CoO$_{2}$ is quasi-2D metal with the a$_{1g}$ hole character dominant at
E$_{F}$. With the intercalation of water, the a$_{1g}$ band becomes more 
narrow and system is near the broader of metal-SC transition.  
We suggest the $Gossamer$ mechanism is responsible for the 
SC in Na$_{0.35}$CoO$_{2}$$\cdot$mH$_{2}$O.

   The CoO$_{6}$ octahedra in Na$_{x}$CoO$_{2}$ are connected via 
sharing their edges forming hexagonal CoO$_{2}$ layers, Na$^{+}$ ions 
are randomly and homogeneously distributed between CoO$_{2}$ layers, 
the intercalation of water molecules into the Co and Na$^{+}$ layers 
does not change the triangular symmetry of the CoO$_{2}$ layers in 
Na$_{x}$CoO$_{2}$$\cdot$mH$_{2}$O.
Under the octahedral O$_{h}$ crystalline electric field, the 3d orbits 
splits into lower threefold degenerate t$_{2g}$ and higher twofold 
degenerate E$_{g}$ levels with the separation of 10$Dq$. The
low spin configuration of Co \cite{s4} showed that only t$_{2g}$ orbitals
are the most relevant to the low energy process in the compounds.
In the coordinate system that the z-axis coincides with the Co-O
bond of the CoO$_{6}$ octahedra, the three wavefunctions of the t$_{2g}$ 
orbitals are: $|a_{1g}>$=\( (d_{x'y'}+d_{y'z'}+d_{z'x'})/\sqrt{3}
\),  $|e_{1g}>$= \( (d_{x'y'}-d_{y'z'})/\sqrt{2}  \) and $|e_{2g}>$=\( 
(-2d_{x'z'}+d_{y'z'}+d_{x'y'})/\sqrt{6} \). 
For the D$_{6h}$ point group symmetry in Na$_{x}$CoO$_{2}$, it is more 
convenient to choose the threefold rotation axis as the z-axis, thus the 
basis wavefunctions of the t$_{2g}$ orbitals are rewritten as: 
$|1>$ =$|a_{1g}>$=\( d_{3z^{2}-r^{2}} \), $|2>$ = $|e_{1g}>$
= \( \sqrt{1/3}(\sqrt{2}d_{xy}+ d_{yz})  \) and  $|3>$ = $|e_{2g}>$= \( 
\sqrt{1/3} (-\sqrt{2}d_{x^{2}-y^{2}}+ d_{xz})  \). 
The trigonal crystalline field in Na$_{x}$CoO$_{2}$ further splits 
t$_{2g}$ orbits as:
 \(  V_{t} (2d^{\dag}_{i1\sigma}d_{i1\sigma} -
         d^{\dag}_{i2\sigma}d_{i2\sigma}-d^{\dag}_{i3\sigma}d_{i3\sigma})
\)
which lifts the a$_{1g}$ orbit 3V$_{t}$ higher than the twofold degenerate
e$_{1g}$ and e$_{2g}$ orbits; the trigonal crystal splitting 3V$_{t}$ 
is about 0.2 eV, estimated from the separation of a$_{1g}$ band to e$_{g}$ 
bands in the following calculation. We will show that in
Na$_{x}$CoO$_{2}$, the e$_{1g}$ and e$_{2g}$ orbitals are almost 
filled, leaving a$_{1g}$ orbit be active, this indicates the orbital 
angular momentum of the 3d electrons in cobalt ions
vanishes, $<L>$=0, therefore the magnetic moment and susceptibility
completely contribute from the spins of 3d electrons. In the 
following we will show that the mixture among the three t$_{2g}$ bands is
weak, in the contrast to vanadium oxide V$_{2}$O$_{3}$, this also excludes
the influence of the orbital degree of freedom on the ground state.

  The crystal structure of Na$_{0.35}$CoO$_{2}$ has already exhibited 
strong 2D character before oxidization, the ratio of the 
lattice constant c/a $\approx$ 3.75, is much large than that of ideal 
three-dimension hexagonal close packed structure. 
The quasi-two-dimension character in the electronic structures of CoO$_{2}$ 
layer is seen clearly from the flat a$_{1g}$ band near FS in Fig.1. 
The electronic structures are obtained by employing the
FP-LMTO method \cite{s17} in the LDA. To explore the realistic situation
with water molecules inserted between Na$^{+}$ and CoO$_{2}$ layers,
we study the evolution of the band structures with increasing
c/a ratio from 3.75, corresponding to Na$_{0.35}$CoO$_{2}$,  4.5, 5.75 and 
up to 6.95, corresponding to Na$_{0.35}$CoO$_{2}$$\cdot$mH$_{2}$O.
\begin{figure}[tbh]
%\vskip 0.50cm
\epsfysize=2.8in\epsfxsize=3in\epsfbox{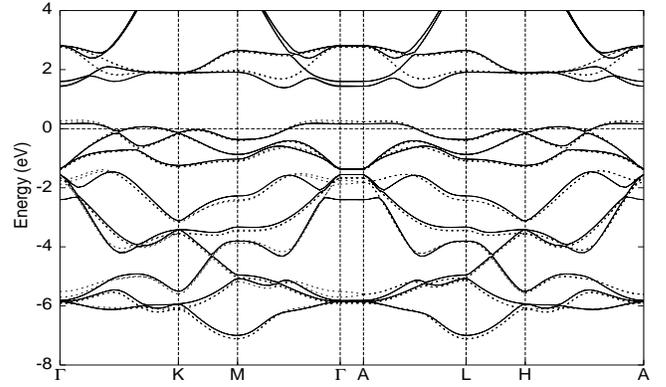}
%\vbox{\epsfxsize=5.0 cm,\epsfbox{fig1}}
\vskip -0.8cm
\caption{
      Band structures of Na$_{0.35}$CoO$_{2}$ within the LDA scheme for
      different c/a ratios. Solid line denotes c/a=6.95, dash line c/a=3.75.
}
\label{fig:fig1}
\end{figure}
The e$_{1g}$ and e$_{2g}$ bands lie below E$_{F}$, the
a$_{1g}$ band crosses the FS and contributes the major component near FS,
this agrees to our crystalline field analysis. Obvious shrink of the
a$_{1g}$ bandwidth is observed in Fig.1 when lattice ratio c/a varies 
from 3.75 to 6.95, indicating that the localization
character of the a$_{1g}$ electrons becomes more significant after the
intercalation of water molecules. In
Fig.1, the bandwidth of the a$_{1g}$ electrons, W = 1.8 eV, giving rise 
to the effective hopping integral between cobalt ions, $|t^{11}_{eff}|$ 
$\approx$ 0.12 eV, which is larger than that in NaCoO$_{2}$ \cite{s8}.
Due to being quasi-2D metal, the band structures near E$_{F}$ 
do not change critically within the LDA+U scheme.

  We further investigate how the density of states (DOS) evolves with 
the doping concentration of Na in Na$_{x}$CoO$_{2}$, from x=0, 0.35, 0.5 to 
0.75 within the usual LDA scheme. The DOS of "virtual" parent phase CoO$_{2}$, 
which has the same structure to Na$_{0.35}$CoO$_{2}$, is also presented
for comparison. It is found that all of the compounds are 
metallic, even for CoO$_{2}$. 
With the doping of Na, the extra electrons from Na are introduced in the 
a$_{1g}$ orbit of Co ions, different from cuprates that the carriers are 
located in the oxygen atom in the Cu-O chain outside the CuO$_{2}$ layers, 
therefore no Zhang-Rice singlet can form in doped 2D cobalt oxides. 
The E$_{g}$ orbitals separate from the t$_{2g}$ 
orbits about 1.7 eV for Na$_{0.35}$CoO$_{2}$, giving rise to the octahedral 
crystalline field splitting 10$Dq$= 1.7 eV. 
Generally the e$_{1g}$ and e$_{2g}$ orbits are almost filled with
the occupation
numbers n$_{e1g}$=n$_{e2g}$ $\approx$ 2.0. The DOS of the 3d electrons at 
E$_{F}$ in Na$_{x}$CoO$_{2}$ is 6.2 states/eV for x=0.35, 3.0 states/eV 
for x=0.5 and 2.8 states/eV for x= 0.75, respectively. 
The DOS of oxygens near E$_{F}$ changes considerably, from 3.8 states/eV
at x=0 to 1.5 states/eV at x=0.75.

    In the LDA scheme the DOS in CoO$_{2}$ at E$_{F}$ is as high as 8.1
states/eV, however, one would expect the ``parent'' compound CoO$_{2}$ with
half-filled a$_{1g}$ orbit be an insulator. Obviously the LDA scheme
underestimates the role of the Coulomb interaction in this system.
The LDA+U scheme \cite{s18} is thus employed to
correct the on-site Coulomb correlation of Co 3d electrons in CoO$_{2}$.
Within the LDA+U scheme with U=5.0 eV, the charge transfer energy between Co 
3d and O 2p orbits in CoO$_{2}$ is $\Delta$ =$\epsilon_{d}-\epsilon_{p}$=3.2 
eV; the bare on-site Coulomb interaction of the Co 3d electrons,
U$_{dd}$ =7.5 eV, which is smaller than about 10 eV of Cu in parent cuprates 
La$_{2}$CuO$_{4}$; $\Delta$ $<<$ U$_{dd}$ implies that the ``parent''
phase CoO$_{2}$ is a charge transfer insulator with energy gap E$_{g}$=1.2
eV. The 3d DOS decreases to zero at E$_{F}$, in contrast with the LDA result
that the system is metallic with very large DOS. We notice that the oxygen
DOS is almost not affected and is dominant over the 3d DOS below E$_{F}$,
as seen the dot line in Fig.2a. In CoO$_{2}$, 
the magnetic moment of cobalt ion is 1.09 $\mu_{B}$, the moment of oxygen
ion is -0.06 $\mu_{B}$, leading to a net moment of 1.03 $\mu_{B}$. 
The planar coupling between spins in CoO$_{2}$ is antiferromagnetic, J =2.9
meV, 
however long range order does not establish due to the spin fluctuation 
arising from the geometric frustration in triangular CoO$_{2}$ layers.
The intraatomic Hund's rule coupling J$_{H}$ of Co in CoO$_{2}$ is similar 
to that in LaCoO$_{3}$ \cite{s19}, smaller than 0.7 eV. These
intraatomic parameters are also applied for doped Na$_{x}$CoO$_{2}$.

    Within the LDA+U scheme, the DOS for different doped compounds
Na$_{x}$CoO$_{2}$ are shown in Fig.2 (b) to (c). The DOS
near FS changes critically for doped compounds we studied (x=0.35, 0.5, 
0.75), in comparison with the LDA results. Though all of the
CoO$_{2}$ layers are metallic, the a$_{1g}$ bands near E$_{F}$ have long
tails, and the DOS at E$_{F}$ is considerably reduced to 1.6, 1.3 and
1.2 states/eV for x=0.35, 0.5, and 0.75, respectively. With the doping, 
the extra
electrons are filled in empty a$^{\downarrow}_{1g}$ band, due to the Coulomb
interaction, the filled  a$^{\uparrow}_{1g}$ band shifts to deep position, 
leaving a$^{\downarrow}_{1g}$ band around E$_{F}$ and smaller the 3d DOS. 
%%
%% The DOS of oxygen 2p bands, as shown in the dot lines in Fig.2, is very
%% small at E$_{F}$.
%%
The magnetic moment of cobalt ion is 0.57 $\mu_{B}$ for x=0.35, 
the moment of oxygen is 0.03  $\mu_{B}$, giving rise to net moment 0.6
$\mu_{B}$; and we find the favorable coupling between Co spins is
ferromagnetic.
The DOS at E$_{F}$ is not large enough to account for the unusual
large thermoelectric power observed in Na$_{x}$CoO$_{2}$, after taking into
account the electron correlation between 3d electrons. One possibility is
the spin entropy of 3d electrons in Na$_{x}$CoO$_{2}$ \cite{s11}.
\begin{figure}[tbh]
%\vskip 0.50cm
\hskip 0.30cm
\epsfysize=3.2in\epsfxsize=3.0in\epsfbox{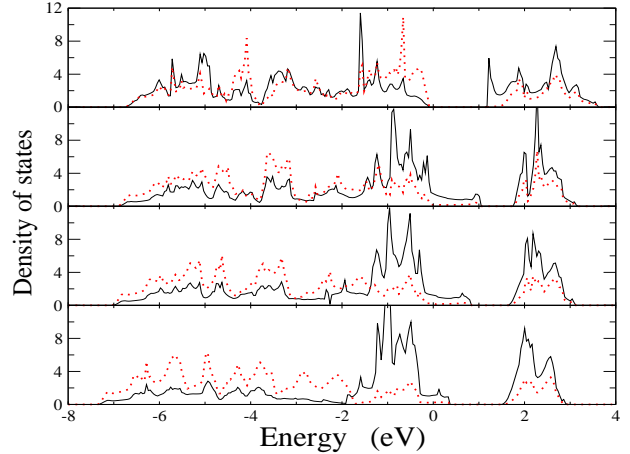}
%\vbox{\epsfxsize=5.0 cm,\epsfbox{fig2}}
\vskip -0.80cm
\caption{
         Dependence of density-of-states in Na$_{x}$CoO$_{2}$ on
         doping concentration within the LDA+U scheme, subsequently, (a) x=0,
         (b) x=0.35, (c) x=0.5, and (d) x=0.75. Solid lines denote DOS of 
         Co 3d orbitals, dot lines DOS of O 2p orbitals.
}
\label{fig:fig2}
\end{figure}
In the LDA scheme, a large and a small FS are obtained in half-doped
compound, Na$_{0.5}$CoO$_{2}$, as the same to Ref.[8]. The large FS arises 
from the
a$_{1g}$ orbit, while small FS pocket from the mixed a$_{1g}$ and e$_{g}$
component of cobalt ions. 
In the LDA +U scheme, the inclusion of 
the correction to the on-site Coulomb interaction repulses the occupied
e$_{g}$ electrons to lower energy, diminishing the contribution of these
electrons to the small FS, thus the small FS disappears.

   We estimate the hopping integrals between 3d orbitals from the $pd$
hybridization between oxygen and cobalt in the CoO$_{2}$ plane. It is found
the Co-O overlap integrals are: t$^{\sigma}_{pd}$ = -1.4 eV, 
t$^{\pi}_{pd}$ = 0.7 eV. The direct hopping integrals between cobalts, 
$dd\sigma$, etc, are very small, since the Co ion radius is small.
From which and considering the $pd$ hybridization effect and the nearly
90$^{\circ}$ angle of the Co-O-Co bridge, we find to the second-order 
approximation, the orbital-diagonal hopping integrals between 
Co 3d orbitals along x-axis are renormalized to: t$_{11}$ $\approx$ -0.09 eV, t$_{22}$
=t$_{33}$ $\approx$ 0.17 eV, respectively; and the mixing hopping
integral t$_{12}$ is very small, about 0.01 eV. The value of 
the hopping integral between a$_{1g}$ orbitals, t$_{11}$, is comparable with 
that from the a$_{1g}$ bandwidth. Intuitively the trigonal splitting V$_{t}$
is so small that three t$_{2g}$ bands, a$_{g1}$, e$_{g1}$ and e$_{g2}$ are
almost degenerate, the orbital degree of freedom of t$_{2g}$ is expected to
play important role in the low energy process. However, the obtained 
small hopping integrals between orbitals suggests the mixing between
orbitals is negligible and the orbital
degree of freedom does not affect the ground state and excited properties in 
quasi-2D cobalt oxides.

\begin{table}[b]
\caption{ Electronic parameters in CoO$_{2}$ \& Na$_{0.35}$CoO$_{2}$ 
(eV)} 
\label{momreal}
%\begin{ruledtabular}
\begin{tabular}{ccccccccccc}
U$_{\text{dd}}$ & J &  E$_{\text{g}}$  & t$^{\pi}_{\text{pd}}$ & 
t$^{\sigma}_{\text{pd}}$  & 10Dq & $\Delta$ & 3V$_{\text{t}}$ &  W &\\
\hline 7.5 & 0.0029 & 1.2 & -1.40 & 0.70 & 1.7 & 3.2 &  0.2 & 1.80 & 
\end{tabular}
%\end{ruledtabular}
\end{table}

%%%
%%%  discuss the role of Coulomb correlation
%%%
As to the electron correlation in 2D cobalt oxides, we note that 
the present experimental data and theoretical results support two distinct 
pictures, one is small Coulomb correlation scenario:  Na$_{x}$CoO$_{2}$
behaves good metallic conduction, and low spin configuration allows
double occupation in 3d orbits; On the other hand, Co ion exhibits
well-defined magnetic moment,
the magnetic susceptibility of Na$_{x}$CoO$_{2}$ exhibits Curie-Weiss
law with finite {\it Neel} temperature $\Theta$ in high temperature 
regime, {\it etc}, supporting a strong correlation scenario. 
In the ``parent'' phase CoO$_{2}$, taking into account the correction 
from the renormalized parameters F$_{2}$ and F$_{4}$ of the 3d multipolar
interaction \cite{s20}, the the bare on-site Coulomb interaction U$_{dd}$
is renormalized to $\tilde{U}_{dd}$ $\approx$ 4.0 eV, suggesting an intermediate
correlated system in comparison with the a$_{1g}$ bandwidth W=1.8 eV.
In the metallic phases Na$_{x}$CoO$_{2}$, due to the screening effect from
the many-body interaction between 3d electrons, the effective Coulomb
strength is further reduced to 
U$_{r}$=$\tilde{U}_{dd}$/(1+$\tilde{U}_{dd}
$$\Lambda$) \cite{s21}, about 2.3 eV in Na$_{0.35}$CoO$_{2}$, where 
$\Lambda$
is the renormalized parameter depending on the conduction band.
%%%
The Wilson ratio deduced from experimental data also supports an
intermediate correlated system: the magnetic susceptibility $\chi$
\cite{s22} and the linear coefficient of the specific heat \cite{s23}
gives rise to Wilson ratio $R$=$(\gamma/k^{2}_{B}{\pi}^{2})/(\chi/
\mu^{2}_{eff}$) =0.62 for $x$=0.5, lying between that of strongly 
correlated system, R=2, and  that of weak correlated Fermi liquid, 
R=1.
Therefore 2D cobalt oxides poss both weak-correlated itinerant and 
strong-correlated localized characters, the SC in 
Na$_{0.35}$CoO$_{2}$$\cdot$mH$_{2}$O seems not arise from the 
instability of the strongly correlated RVB metal, the RVB scenario is not 
applicable.
%%%

   In this study  we try to unveil the role of water molecules to the novel 
SC in  hydrated cobalt oxides. The intercalation of water 
molecules into CoO$_{2}$ and Na layers is simulated by stretching the 
c-axis, since no configuration information about the water molecules in
Na$_{0.35}$CoO$_{2}$$\cdot$mH$_{2}$O is available. 
This may not completely reproduce the realistic crystalline environment 
of the CoO$_{6}$ octahedra. 
The role of intercalation of water lies in three aspects. First of all, 
it enhances the localization of the a$_{1g}$ electrons, pushes the 2D 
CoO$_{2}$ layers to the broader of metal-insulator transition, similar 
to the effect of doping on cuprates, favors the instability of 
metal-SC transition. Second, 
with the intercalation of water molecules, 
the hydrogens of water molecules approaching CoO$_{6}$ octahedron 
tend to form hydrogen bond with the vertical oxygen of the CoO$_{6}$
octahedra, the negative charge center in O$^{2-}$ shifts to H$^{+}$, thus
we have such possibility that a fraction of holes may introduced in 
filled 2p band of oxygen, the spatial separation between the $a_{1g}$ 
electrons and the $2p$ holes blocks the their combination. Third, 
though there exists strong electron-phonon (e-ph) coupling in 2D 
CoO$_{2}$ layers in Bi$_{2}$Sr$_{2}$Co$_{2}$O$_{y}$, one expects the
e-ph coupling is not weak in Na-doped CoO$_{2}$, the SC
transition temperature would be high if considering the large DOS at 
E$_{F}$ in the absence of Coulomb correlation; the 
elongation of c-axis in Na$_{x}$CoO$_{2}$ weakens the vibration frequency
of Co-O bond along c-axis, the vibration mode along c-axis is soften, 
this is not favorable of the e-ph coupling mediated pairing mechanism, 
but favors to other mechanism. The isotope effect experiment of the 
oxygens on CoO$_{2}$ layer is expected to confirm this result.

 In 2D triangular CoO$_{2}$ layers the intermediate electronic correlation
allowing double occupation excludes RVB SC, whileas, 
the antiferromagnetic spin fluctuation and the frustration effect are 
important, the antiferromagnetic long range order is suppressed. 
For the correlated electron system with intermediate Coulomb strength, 
like the present compound, Laughlin and Zhang \cite{s24} suggested in the new
quasiparticle basis allowing double occupation in the electronic orbits,
the electrons experience a powerful attracting force, providing   
a pairing mechanism, named {\it Gossamer} SC. 
In such a {\it Gossamer} SC, the superfluid carrier density is 
very thin, it is naturally expected the transition temperature T$_{C}$ 
is low.

   In summary, two-dimensional triangular CoO$_{2}$ layer is intermediate 
correlated electron system, Na$_{0.35}$Co$_{2}$ is quasi-2D metal with dominant
a$_{1g}$ hole character. The intercalation of water enhances the 
localization character of a$_{1g}$ carriers, the superconductivity arises from 
{\it Gossamer} pairing mediated by the spin fluctuations.
\\

%\acknowledgments
   One of authors L-J. Zou thanks M. Fabrizio. 
Financial supports from the NNSF of China No.10174084 and from the Chinese
Academy of Sciences (CAS) are appreciated.


\begin{references}

\bibitem{s1} P. W. Anderson, {\it Science} {\bf 235}, 1196 (1987);
             G. Baskaran, Z. Zou and P. W. Anderson, {\it Solid State Commun.}, 
             {\bf 63}, 973 (1987).
\bibitem{s2} I. Terasaki, Y. Sasago, and K. Uchinokura, {\it Phys. Rev.} 
             {\bf B 56}, 12685 (1997).
\bibitem{s3} T. Kawata, {\it et al}. {\it Phys. Rev.} {\bf B 60}, 10584 (1999).
\bibitem{s4} S. Nakamura and A. Ozawa, {\it J. Phys. Soc. Jpn} {\bf 68}, 3746 (1999)
             and some references therein.
\bibitem{s5} T. Valla, {\it et al}., {\it Nature}, {\bf 417}, 627 (2002).
\bibitem{s6} T. Tanaka, S. Nakamura and S. T. Iida, {\it Jpn. J. Appl. Phys.}
             {\bf 33}, 581 (1994).
\bibitem{s7} T. Itoh, {\it et al}.  {\it LANL Preprint}, condmat/0304377.
\bibitem{s8} R. P. Singh, {\it Phys. Rev.} {\bf B 61}, 13397 (1999).
\bibitem{s9} W. Koshibae, K. Tsutsui and S. Maekawa, {\it Phys. Rev.} {\bf B 62}, 
              6869 (2000); {\it J. Mag. Mag. Mat.}, {\bf 268}, 216 (2003).
\bibitem{s10} T. Motohashi, {\it et al}., {\it Phys. Rev.} {\bf B 67}, 064406 (2003).
\bibitem{s11} Yayu Wang, {\it et al}., {\it Nature}, {\bf 423}, 425 (2003).
\bibitem{s12} K. Takada, {\it et al}., {\it Nature} {\bf 422}, 53 (2003).
\bibitem{s13} B. Lorenz, {\it et al}., {\it LANL Preprint}, condmat/0304537.
\bibitem{s14} F. C. Chou, {\it et al}., {\it LANL Preprint}, condmat/0306659.
\bibitem{s15} G. Baskaran, {\it LANL Preprint}, condmat/0303649.
\bibitem{s16} Q.-H. Wang D.-H. Lee, and P. A. Lee, {\it LANL Preprint}, 
              condmat/0304377.
\bibitem{s17} S. Y. Savrasov, {\it Phys. Rev.} {\bf B 54}, 16470 (1996).
\bibitem{s18} A. I. Liechtenstein, V. I. Anisimov, and J. Zaanen, {\it Phys. Rev.} 
              {\bf B 52}, 5467 (1995).
\bibitem{s19} C. Zobel, {\it et al}.,{\it Phys. Rev.} {\bf B 66}, 020402 (2002).
\bibitem{s20} S. Sugano, Y. Tanabe and H. Kamimura, {\it Multiplets of Transition-Metal 
              Ions in Crystal}, Academic Press, New Rork (1970);
              A. Tanaka, {\it LANL Preprint}, condmat/0201407.
\bibitem{s21} L. Chen, {\it et al}. {\it Phys. Rev. Lett.}, {\bf 66}, 369 (1991)
\bibitem{s22} F. Rivadulla, J. S. Zhou, and J. B. Goodenough, {\it LANL Preprint}, 
              condmat/0302426.
\bibitem{s23} Y. Ando, {\it et al}., {\it Phys. Rev.} {\bf B 60}, 10580 (1999).
\bibitem{s24} Laughlin, {\it Phys. Rev.} {\bf B 61}, 11506 (2000); F. C. Zhang, 
              {\it Phys. Rev. Lett.} {\bf B 78}, 11506 (2003).

\end{references}
\end{document}